\documentclass[english,nofootinbib, pra,twocolumn,superscriptaddress,english]{revtex4-1}
\usepackage[T1]{fontenc}
\usepackage[latin9]{inputenc}
\usepackage{color}
\usepackage{babel}
\usepackage{mathtools}
\usepackage{amsmath}
\usepackage{amssymb}
\usepackage{graphicx}
 \usepackage[colorlinks,linkcolor=blue,anchorcolor=blue,citecolor=blue,urlcolor=blue]{hyperref}
 
\usepackage{amsfonts}
\usepackage{bm}
\usepackage{makeidx}

\makeatletter
\usepackage{bbm}

\makeatother

\begin{document}
\title{Efficient measurement of the time-dependent cavity field through compressed sensing}
\author{Fang Zhao}
\affiliation{Center for Quantum Technology Research and Key Laboratory of Advanced
Optoelectronic Quantum Architecture and Measurements (MOE), School
of Physics, Beijing Institute of Technology, Beijing 100081, China}
\affiliation{China Academy of Engineering Physics, Beijing 100088, People's Republic
of China}
\author{Qing Zhao}
\affiliation{Center for Quantum Technology Research and Key Laboratory of Advanced
Optoelectronic Quantum Architecture and Measurements (MOE), School
of Physics, Beijing Institute of Technology, Beijing 100081, China}
\author{Dazhi Xu}
\email{dzxu@bit.edu.cn}

\affiliation{Center for Quantum Technology Research and Key Laboratory of Advanced
Optoelectronic Quantum Architecture and Measurements (MOE), School
of Physics, Beijing Institute of Technology, Beijing 100081, China}
\begin{abstract}
We propose a method based on compressed sensing (CS) to measure the evolution processes of the states of a driven cavity quantum electrodynamics system. In precisely reconstructing the coherent cavity field amplitudes, we have to prepare the same states repetitively and each time perform one measurement with short sampling intervals considering the quantum nature of measurement and the Nyquist--Shannon sampling theorem. However, with the help of CS, the number of measurements can be exponentially reduced without loss of the recovery accuracy. We use largely detuned atoms and control their interactions with the cavity field to modulate coherent state amplitudes according to the scheme encoded in the sensing matrix. The simulation results show
that the CS method efficiently recovers the amplitudes of the coherent
cavity field even in the presence of noise. 
\end{abstract}
\maketitle

\section{Introduction}

A measurement usually has destructive consequences for the quantum
state, which prevents repetitive probing of the quantum system.
Even though the quantum non-demolition measurement allows the measurement of an observable without destroying the system~\citep{1980.Thorne}, this method has limitations. It generally fails in the case of frequent
measurements owing to the quantum Zeno effect~\citep{1977.Sudarshan,1990.Wineland}. A typical method of recording the time evolution of a quantum system is to prepare an ensemble of identical quantum systems and perform
only one measurement for each copy at a different sampling instant.
However, this approach requires enormous discrete samplings to reconstruct the original continuous-time dynamic signals. The minimal sampling rate of this approach is given by the Nyquist--Shannon sampling theorem~\citep{1928.Nyquist,1949.Shannon}.

Fortunately, the theory of compressed sensing (CS) developed by mathematicians nearly two decades ago~\citep{2006.Taoitr,2006.Tao,2006.Donoho} has
proven to be able to exactly recover sparse signals with only
a small set of linear random measurements. As sparsity is a general
property of almost all meaningful signals, CS techniques have
been widely applied in various areas, such as medical imaging~\citep{2007.Pauly}, radar~\citep{2010.Cetin,2019.Zhao}, and geophysics~\citep{2007.Herrmann}. Recently, accompanying the fast development of quantum sensing, many attempts have been made to use CS techniques to improve the speed of quantum measurements. Most of these quantum CS schemes concern time-independent signals; e.g., quantum imaging~\citep{2009.Silberberg,2009.Silberberg4a6,2011.Boydefx},
quantum tomography~\citep{2010.Eisert4qn, 2010.Liu, 2011.White, 2015.Joynt,2017.Eiserth38}, and two-dimensional spectroscopy~\citep{2013.Kubarych,2017.Brixner}. 

To measure time-dependent signals in a compressive way, one needs
to control the dynamics of the quantum system effectively. In an interesting work~\citep{2013.Cappellaro}, the authors proposed a method of compressively measuring the time-varying magnetic field via a single spin, where a sequence of $\pi$-pulses flips the spin according to the requirements of the CS theory. However, the spin in Ref.~\citep{2013.Cappellaro} is along the magnetic field; thus, the Hamiltonians at different times commute with each other ($\left[H(t),H(t^{\prime})\right]=0$), which only covers a particular type of quantum dynamics. On the basis of a cavity quantum electrodynamics (QED) system, we show that the idea of the compressive measurement of the time-dependent signals can be generalized to a quantum system with the property $\left[H(t),H(t^{\prime})\right]\neq0$ ($t\neq t^{\prime}$) in this work.

The cavity QED system is an essential platform for quantum computation~\citep{2010.O'Brien} and quantum sensing~\citep{2019.Lehnert,2020.Rey}.
We employ CS techniques to continuously measure the state of the
cavity field driven by a time-varying classical source. Our purpose
is to reconstruct the coherent cavity amplitude with high accuracy
while exponentially reducing the number of measurements. To this
end, we send sequences of largely detuned atoms through the cavity at
randomly chosen instants during the evolution processes, such that
the cavity field can be controlled according to the assigned sensing matrix. Each measurement is performed at the end of one complete time evolution process. With only a small set of under-sampled measurement
results, the convex optimization algorithm can successfully recover
the coherent amplitude even in the presence of white noise. Our approach
expands the application range of CS techniques in the measurements
of quantum systems.

The paper is arranged as follows. Section~II revisits the theory
of CS. Section~III briefly introduces the driven cavity
QED system and the Nyquist--Shannon sampling method for the cavity
field. Section~IV shows how to apply CS techniques to the cavity
QED system. Section~V presents the recovery results and the error analysis. Section~VI summarizes our results and discusses the
possible future work. 

\section{Compressed Sensing\label{sec:II}}

The Nyquist--Shannon theorem states that to exactly recover a continuous-time
signal, the sampling frequency should be at least twice the signal's
highest frequency~\citep{1928.Nyquist,1949.Shannon}. If the signal
has a sparse representation in a set of known bases, the CS theory
allows a much lower sampling frequency than that required by the Nyquist--Shannon theorem. Before applying the CS techniques to the cavity QED system, we briefly introduce the basic knowledge of CS theory in this
section. In the linear measurement models, the signal to be measured
can be modeled using an $N$-dimensional vector $\vec{x}\in\mathbb{R}^{N}$,
and the measurement is formulated as an $M\times N$ matrix $\mathcal{A}$
acting on $\vec{x}$:
\begin{equation}
\vec{y}=\mathcal{A}\cdot\vec{x},\label{eq:CS}
\end{equation}
where the $M$-dimensional vector $\vec{y}\in\mathbb{R}^{M}$ is the measurement results and $M$ is the number of measurements. If $M=N$ and $\mathcal{A}$ is a full-rank matrix, the signal can be uniquely
and exactly reconstructed according to $\vec{x}=\mathcal{A}^{-1}\cdot\vec{y}$.
However, we suppose the number of measurements $M$ is smaller than the
dimension $N$ of the signal. In that case, there are infinite numbers
of $\vec{x}$ satisfying Eq.~(\ref{eq:CS}), and it is thus difficult to tell which one is the actual signal without further information. Fortunately,
real-world signals are generally structured, and sparsity is one of
the essential prior characteristics of the signals. A signal is called
$S$-sparse if it has no more than $S$ ($S\ll N$) non-zero components.
The CS technique takes advantage of sparsity to recover the signal
from an appreciable reduced number of samplings. 

Mathematically, theory says if two conditions
are satisfied, (1) $\vec{x}$ is $S$-sparse and (2) $\mathcal{A}$ satisfies the restricted isometry property (RIP) of order $2S$, then the unique $S$-sparse $\vec{x}$ can be exactly recovered by solving the convex optimization problem: 
\begin{equation}
\vec{x}=\arg\min\Vert\vec{x}^{\prime}\Vert_{1}~\text{suject to}~\vec{y}=\mathcal{A}\cdot\vec{x}^{\prime},\label{eq:ConOptim}
\end{equation}
where $\vec{x}^{\prime}$ is the recovered signal and $\Vert\vec{x}^{\prime}\Vert_{1}\coloneqq\sum_{i}\vert x_{i}^{\prime}\vert$
is the $l_{1}$-norm of vector $\vec{x}^{\prime}$. The RIP of order $S$ requires
$\mathcal{A}$ satisfying
\begin{equation}
(1-\delta_{S})\Vert\vec{x}\Vert_{2}^{2}\leq\Vert\mathcal{A}\cdot\vec{x}\Vert_{2}^{2}\leq(1+\delta_{S})\Vert\vec{x}\Vert_{2}^{2}\label{eq:RIP}
\end{equation}
for all $S$-sparse vectors $\vec{x}$. Here, $\Vert\vec{x}\Vert_{2}^{2}$
is the $l_{2}$-norm or Euclidean distance of $\vec{x}$, and $\delta_{S}\in(0,1)$ is an $S$-dependent number~\citep{2006.Taoitr,2008.Wakinfgl}. The RIP of order $2S$ is a sufficient condition of the sensing matrix $\mathcal{A}$ for successfully recovering the unique $S$-sparse signal, which can be intuitively understood as a restriction on $\mathcal{A}$ such that it approximately preserves the distance between any two $S$-sparse vectors. A smaller $\delta_{S}$ corresponds to $\mathcal{A}$ being more similar to an isometry transformation. It is usually difficult to construct a matrix that strictly satisfies the RIP. However, it has been proved that a random matrix can satisfy the RIP with very high probability~\citep{2006.Taoitr},
which makes the application of CS techniques extremely convenient.

There are various recovery algorithms suitable for solving the convex
optimization problem of Eq.~(\ref{eq:ConOptim}). We use the orthogonal
matching pursuit (OMP) algorithm~\citep{1993.Krishnaprasad}, which
is accurate and efficient for the goals of this work. If the signal
is not sparse in the bases in which we perform the measurements, we first
find an orthogonal transformation $\Phi$ such that $\vec{x}_{\Phi}=\Phi\cdot\vec{x}$
is sparse. The transformation $\Phi$ is generally known for a class
of signals in prior. For instance, the images are usually sparse after
wavelet transformation. As a result, we can reformulate the constraint
in the optimization problem of Eq.~(\ref{eq:ConOptim}) as $\vec{y}=\mathcal{A}_{\Phi}\cdot\vec{x}_{\Phi}^{\prime}$,
where the transformed sensing matrix $\mathcal{A}_{\Phi}=\mathcal{A}\cdot\Phi^{-1}$
is a random matrix satisfying the RIP as long as $\mathcal{A}$ is
random. 

Notwithstanding that the goal of CS is to reduce the number of linear measurements $M$ as much as possible, it is evident that $M$ cannot be made arbitrarily small to extract sufficient information of the signal. If the sensing matrix $\mathcal{A}$ satisfies the RIP of order $2S$ with $\delta_{2S}\in(0,\frac{1}{2}]$, then the lower bound of $M$ is given by
\begin{equation}
M\geq C\cdot S\log_{2}(\frac{N}{S}),\label{eq:Mbound}
\end{equation}
where $C$ is a constant~\citep{2008.Wakinfgl}. This theorem shows
that CS can reduce the number of linear measurements from the
order of $\mathcal{O}(N)$ to the order of $\mathcal{O}(\log N)$. 

Furthermore, as the idea of CS is to extract the information of only
the most significant elements of the signal, CS techniques naturally
resist low noise in the signal. In the presence of low noise with strength $\epsilon$ in the signal, we can generalize
the convex optimization problem of Eq.~(\ref{eq:ConOptim}) as
\begin{equation}
\vec{x}=\arg\min\Vert\vec{x}^{\prime}\Vert_{1}~\text{suject to}~\Vert\vec{y}-\mathcal{A}\cdot\vec{x}^{\prime}\Vert_{2}^{2}\leq\epsilon.\label{eq:CSrec}
\end{equation}
The noise resistance property is another advantage of the CS techniques
for our purpose of reconstructing the time evolution of the cavity field. 

\section{Nyquist--Shannon sampling of the driven cavity field\label{sec:III}}

Techniques used for the instantaneous measurement of the coherent field
or the average photon number inside the microwave cavity have been
well developed; e.g., homodyne detection~\citep{2009.Raymer} and
quantum non-demolition measurement~\citep{1992.Zagury,1999.Haroche}.
Continuous probing of the varying cavity field not only provides
more information about the dynamics of the quantum system but also
is helpful to the quantum precision measurement of external parameters.
However, an efficient method of continuously interrogating the cavity
QED system is still lacking owing to the fragility of the quantum system.
To this end, we adopt CS techniques to accelerate the sampling processes.
For preparation, this section introduces the driven cavity system
with the usual continuous measurement procedure without CS techniques.

The system that we are interested in is a microwave cavity driven by a
classical time-dependent radio-frequency field, the Hamiltonian of
which in the rotating frame is
\begin{equation}
H(t)=if(t)(e^{-i\Delta t}a^{\dagger}-e^{i\Delta t}a).\label{eq:Ht}
\end{equation}
Here, $a(a^{\dagger})$ is the annihilation (creation) operator of
the cavity field, and $\Delta=\omega_{\text{d}}-\omega_{0}$ is the detuning between the driving field frequency $\omega_{\text{d}}$ and the cavity mode $\omega_{0}$. The amplitude of the driving field $f(t)$ is
assumed to be a real function of time. The evolution operator corresponding
to $H(t)$ is given by
\begin{equation}
U(t,t_{0})=e^{i\phi(t,t_{0})}D[\alpha(t,t_{0})],\label{eq:UD}
\end{equation}
where $\exp[i\phi(t,t_{0})]$ is a global phase factor with
\begin{equation}
\phi(t,t_{0})=\int_{t_{0}}^{t}ds\int_{t_{0}}^{s}ds^{\prime}f(s)f(s^{\prime})\sin\left[\Delta(s-s^{\prime})\right]\label{eq:phi}
\end{equation}
and $D[\alpha]\equiv\exp[\alpha a^{\dagger}-\alpha^{\ast}a]$ is the
displacement operator with the time-dependent parameter
\begin{equation}
\alpha(t_{2},t_{1})=\int_{t_{1}}^{t_{2}}dsf(s)e^{-i\Delta s}.\label{eq:alpha}
\end{equation}
It is straightforward to find that $\alpha(t_{2},t_{1})$ satisfies the relation $\alpha(t_{3},t_{1})=\alpha(t_{3},t_{2})+\alpha(t_{2},t_{1})$
with $t_{3}\geq t_{2}\geq t_{1}$. The cavity is assumed to be initially
prepared in the vacuum state $\vert\psi(t_{0})\rangle=\vert0\rangle$;
the average photon number $\bar{n}(t)=\langle\psi(t)\vert a^{\dagger}a\vert\psi(t)\rangle=|\alpha(t,t_{0})|^{2}$
is then a function of time, where $\vert\psi(t)\rangle=U(t,t_{0})\vert\psi(t_{0})\rangle$. 

The coherent state amplitude $\alpha$ of a steady field can be obtained from a balanced homodyne measurement~\citep{2001.Schiller,2009.Raymer}.
However, the continuous measurement of the amplitude of a varying photon field is not easy because the measurements inevitably perturb the quantum state and alter its photon statistics even with the help of the quantum non-demolition measurement~\citep{1992.Ogawa}.
Moreover, the repeated measurement of the cavity field can
freeze the evolution of the cavity state owing to the quantum Zeno effect~\citep{2008.Haroche,2011.Sun}. 

\begin{figure}
\includegraphics[width=9cm]{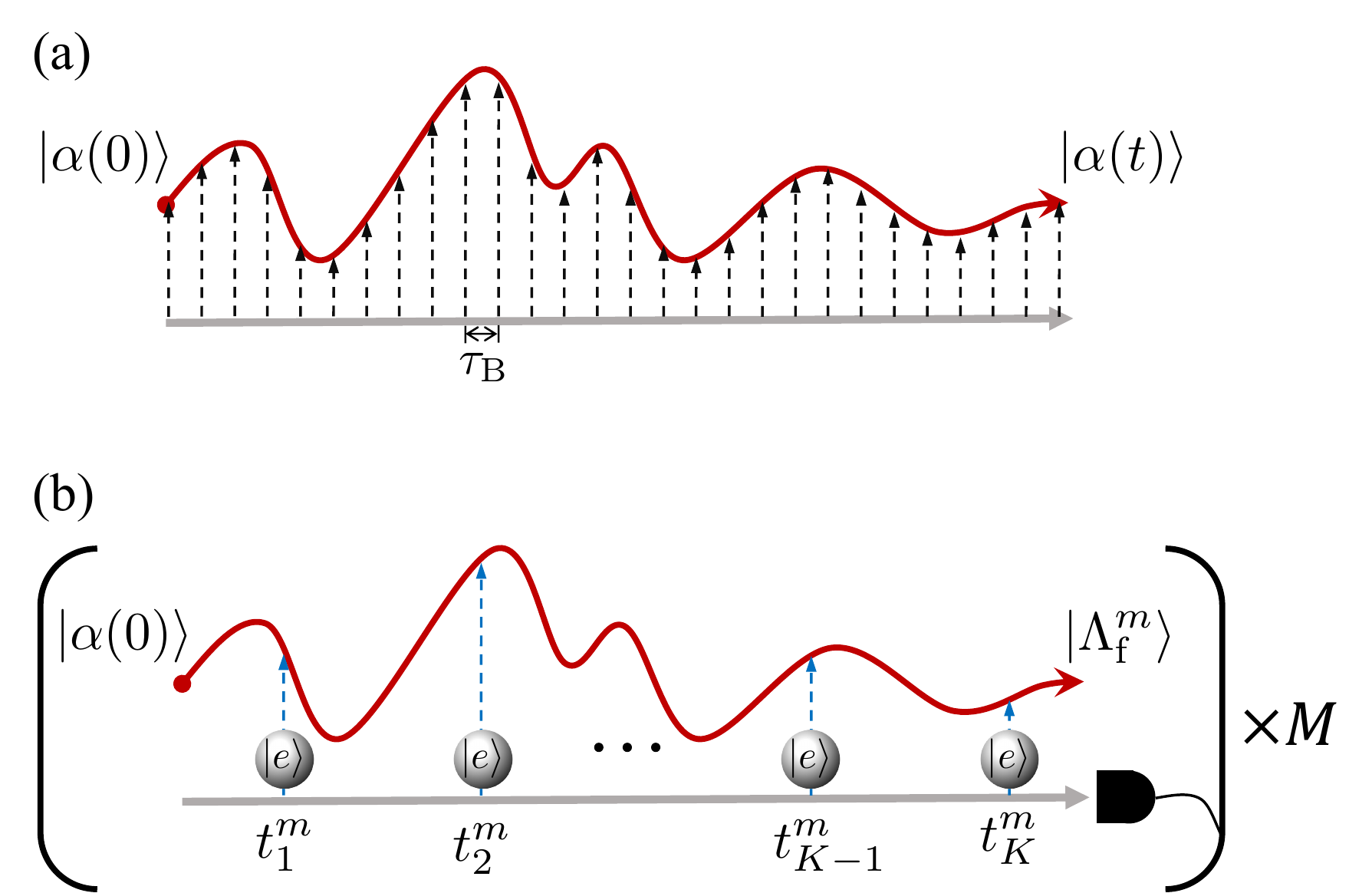}\caption{\label{fig:1}Schematic illustration of the sampling processes of
the state of the cavity field driven by a time-dependent signal: (a)
the Nyquist--Shannon sampling sequence with uniform sampling interval
$\tau_{\text{B}}^{-1}=2B$; (b) the CS sampling scheme
requires $M$ samplings each at the end of one complete evolution process,
during which the phase of the field is altered through interaction with
$K$ largely detuned atoms passing through the cavity at randomly chosen
moments.}
\end{figure}

Therefore, it is impossible to make all the measurements within
one unitary evolution in recovering the continuous signal $\alpha(t,t_{0})$ within an interval $t\in[t_{0},t_{N}]$. Instead, one straightforward approach is to prepare $N$ identical initial states repetitively and each time let the system evolve according to Eq.~(\ref{eq:UD}) until a
different moment $t_{n}\equiv n\tau_{\text{B}}$ $(n=1,2,\dots,N)$
when the measurement is carried out. Here, $\tau_{\text{B}}$ is the
uniform sampling interval, which is chosen as $\tau_{\text{B}}^{-1}\geq2B$
according to the Nyquist--Shannon sampling theorem. The sampling sequence
is illustrated in Fig.~\ref{fig:1}(a). The total number of measurements
is then $N=t_{N}/\tau_{\text{B}}$. The resource consumed in the experiments linearly grows with the number of measurements $N$, which is proportional to the bandwidth of the signal.

In the language of signal processing, the time-domain discretized
coherent state amplitude $\alpha(t,t_{0})$ is represented by an $N$-dimensional
column vector $\vec{\alpha}=(\alpha_{1},\dots,\alpha_{N})^{\text{T}}$
with components $\alpha_{n}\equiv\alpha(t_{n},t_{0})$, which is the
signal that we aim to recover. The above Nyquist--Shannon sampling procedure is a trivial version of linear measurement described by Eq.~(\ref{eq:CS}), where $\vec{\alpha}$ corresponds to the vector $\vec{x}$, $\vec{y}$
is an $N$-dimensional column vector, each component of which is measured
in one experiment. We therefore have $y_{n}=\alpha_{n}$, and $\mathcal{A}$ is a trivial $N\times N$ unit matrix without considering experimental error.

\section{Compressed sampling of the driven cavity field\label{sec:IV}}

The advantage of adopting CS techniques relies on exponentially
reducing the sampling rate in reconstructing the dynamic parameters,
such as $\alpha(t,t_{0})$ and $f(t)$. Then, only $M$ ($M\ll N$)
samplings are needed to construct a low-dimensional column vector
$\vec{y}$. The critical step toward recovering $\vec{\alpha}$ is
the design and realization of the sensing matrix $\mathcal{A}$ with
the experimental feasible quantum control methods. 

Mathematically, each row of the sensing matrix $\vec{\mathcal{A}}^{m}$
$(m=1,2,\dots,M)$ acting on the vector $\vec{\alpha}$ yields a linear
combination of the components of the coherent amplitude $\vec{\alpha}$.
Specifically, in the cavity QED setup, such linear manipulation of
the cavity field can be realized by sending an elaborately designed
sequence of largely detuned two-level atoms through the cavity.
The effective Hamiltonian describing the interaction between the cavity
field and atom is given by 
\begin{equation}
H_{\text{A}}=\xi a^{\dagger}a(\vert e\rangle\langle e\vert-\vert g\rangle\langle g\vert).\label{eq:H_M}
\end{equation}
Here, $|e\rangle$ and $|g\rangle$ are the atomic excited and ground
states, respectively. The effective coupling strength $\xi=g^{2}/(\omega_{0}-\omega_{\text{a}})$
is defined by the Rabi frequency $g$, the cavity mode frequency $\omega_{0}$ and the frequency splitting $\omega_{\text{a}}$ between the atomic
levels. 

Each atom is assumed to interact with the cavity field for a very
short time $\tau_{\text{A}}$ before leaving the cavity. The corresponding
evolution operator is $\exp(-iH_{\text{A}}\tau_{\text{A}})$. We also
assume that $\xi$ is much larger than the average strength of the driven
field $f(t)$ during $\tau_{\text{A}}$; thus, the unitary evolution
$U(t+\tau_{\text{A}},t)$ can be neglected when the atom is interacting
with the cavity field. If the atom is prepared in the initial state
$\vert e\rangle$ and the cavity field in the coherent state $\vert\alpha\rangle$
at the instant that the atom enters the cavity, the state of the atom--cavity system becomes
\begin{equation}
e^{-iH_{\text{A}}\tau_{\text{A}}}\vert\alpha\rangle\otimes\vert e\rangle=\vert\alpha e^{-i\xi\tau_{\text{A}}}\rangle\otimes\vert e\rangle\label{eq.Rphi}
\end{equation}
once the atom leaves the cavity. It is seen that the atom does
not alter the average photon number $\bar{n}=\vert\alpha\vert^{2}$
because the dispersive coupling in Eq.~(\ref{eq:H_M}) conserves the photon number. However, the cavity field acquires a phase factor $e^{-i\xi\tau_{\text{A}}}$
associated with the atomic state. 

Once the atoms are prepared in the excited state $\vert e\rangle$,
they will not evolve because of the dispersive coupling. As a result,
one can only consider the cavity field state, whose evolution operator
is given by a projection on $\vert e\rangle$ as
\begin{equation}
R(\tau_{\text{A}})=\langle e\vert e^{-iH_{\text{A}}\tau_{\text{A}}}\vert e\rangle=e^{-i\xi\tau_{\text{A}}a^{\dagger}a}.\label{eq:R}
\end{equation}
We use this phase-modulation operator $R(\tau_{\text{A}})$ as
a tool with which to manipulate the cavity field according to the arrangement set by the sensing matrix.

\subsection*{Procedures of CS measurements in the cavity QED system}

All the requirements for the CS measurements have
been prepared. We list the main procedures of CS measurements of the cavity field as follows.

(i) Initially prepare the cavity field in the vacuum state $\vert0\rangle$.

(ii) Drive the cavity field using the classical external source $f(t)$
from time $t_{0}$ to $t_{N}$. During the evolution, largely detuned
two-level atoms prepared in the excited state pass through the cavity
field separately at $K$ randomly chosen moments $t_{1},t_{2},\dots,t_{K}$.
All the $K$ moments should be positive integer multiples of $\tau_{\text{B}}$.
The atom--photon interaction time can be tuned using the velocity of the
atom and is set at $\tau_{\text{A}}^{\ast}=\pi/\xi$, during which
the free evolution of the cavity field can be neglected.

(iii) At time $t_{N}$, the coherent field amplitude $\Lambda_{\text{f}}$
is measured via homodyne detection, which completes one compressed
sampling of a single data point. 

(iv) Return to the beginning and repeat steps (i) to (iii) $M$ times. The $M$ measured results $\Lambda_{\text{f}}^{m}$ ($m=1,2,\dots,M$
labels the $m$th measurement) form a column vector $\vec{\Lambda}_{\text{f}}=(\Lambda_{\text{f}}^{1},\Lambda_{\text{f}}^{2},\dots,\Lambda_{\text{f}}^{M})^{\text{T}}$,
which is the signal obtained by compressive sampling.

(v) The time-dependent cavity field amplitude is recovered by the
convex optimization algorithm with the input $\vec{\Lambda}_{\text{f}}$
and the sensing matrix constructed in step (ii).

A schematic illustration of the CS approach is shown in Fig.~\ref{fig:1}
(b).

\subsection*{Sensing matrix}

We now take the $m$th measurement as an example to illustrate how
to construct the sensing matrix $\mathcal{A}$ based on the dynamic
evolution of the cavity state. The initial state of the cavity field
is the vacuum state $|\psi(t_{0})\rangle=\vert0\rangle$, and the
$K$ injected atoms are always in the product state $\vert\mathcal{E}\rangle\equiv\otimes_{j=1}^{K}\vert e\rangle_{j}$.
Before the first atom enters the cavity at the randomly chosen moment
$t_{1}^{m}$, the cavity field evolves into a coherent state 
\begin{equation}
\vert\psi^{m}(t_{1}^{m})\rangle=U(t_{1}^{m},t_{0})\vert\psi(t_{0})\rangle=\vert\alpha(t_{1}^{m},t_{0})\rangle.\label{eq:psit1}
\end{equation}
At $t_{1}^{m}$, the first atom starts interacting with the cavity
field for a short interval $\tau_{\text{A}}^{\ast}$. When the interaction
ends, the sign of the cavity field amplitude is flipped as 
\begin{equation}
R(\tau_{\text{A}}^{\ast})\vert\psi^{m}(t_{1}^{m})\rangle=\vert-\alpha(t_{1}^{m},t_{0})\rangle.\label{eq:UA}
\end{equation}
Here, the role of the excited state atom is akin to the $\pi$-pulse
that flips the spin in the NMR system. As in Ref.~\citep{2013.Cappellaro},
the varying magnetic field is detected using CS techniques, where
the spin is flipped by the $\pi$-pulses serving as a time-domain
modulation of the magnetic field signal. In our work, the excited
state atoms alter the sign of the field amplitude, which is the modulation of the integrated signal of the external parameter $f(t)$ [Eq.~(\ref{eq:alpha})]. We note that $\tau_{\text{A}}^{\ast}=\pi/\xi$
is not a unique option; any value except $\tau_{A}$ equaling integer multiplies of $2\pi/\xi$ will work, and we here consider only the
simplest case. 

$U(t_{j+1}^{m},t_{j}^{m})$ and $R(\tau_{\text{A}}^{\ast})$ alternately
govern the subsequent time evolution according to the randomly generated
time sequence $t_{1}^{m},t_{2}^{m},\dots,t_{K}^{m}$. We can use the
method of induction to derive the expression of the cavity state.
Generally, the state of the cavity field when the $j$th atom leaves
the cavity can be written as
\begin{equation}
|\psi^{m}(t_{j}^{m})\rangle=e^{i\Theta_{j}^{m}}\vert\Lambda_{j}^{m}\rangle,\label{eq:psi_tj}
\end{equation}
where $\Lambda_{j}^{m}$ is the coherent state amplitude at $t_{j}^{m}$ and $\Theta_{j}^{m}$ is an accumulated global phase. During the subsequent evolution from $t_{j}^{m}$ to $t_{j+1}^{m}$, the cavity is first driven by $U(t_{j+1}^{m},t_{j}^{m})$ and then interacts with the
$(j+1)$th atom. Therefore, when the $(j+1)$th atom leaves the cavity,
the state of the cavity field reads
\begin{align}
\vert\psi^{m}(t_{j+1}^{m})\rangle & =R(\tau_{\text{A}}^{\ast})U(t_{j+1}^{m},t_{j}^{m})\vert\psi^{m}(t_{j}^{m})\rangle\nonumber \\
 & =e^{i\Theta_{j+1}^{m}}\vert-[\Lambda_{j}^{m}+\alpha(t_{j+1}^{m},t_{j}^{m})]\rangle,\label{eq:psi_tj+1}
\end{align}
from which it is straightforward to find the recursive relation $\Lambda_{j+1}^{m}=-\Lambda_{j}^{m}-\alpha(t_{j+1}^{m},t_{j}^{m})$.With
the initial value $\Lambda_{1}^{m}=-\alpha(t_{1}^{m},t_{0}^{m})$,
we find
\begin{equation}
\Lambda_{j}^{m}=\sum_{k=1}^{j}(-1)^{j-k+1}\alpha(t_{k}^{m},t_{k-1}^{m}).\label{eq:Lambda_j}
\end{equation}
The accumulated global phase can be derived similarly as $\Theta_{j}^{m}=\sum_{k=1}^{j}\text{Im}\left[\alpha(t_{k}^{m},t_{k-1}^{m})\Lambda_{k-1}^{m\ast}\right]$,
though it is not related to this work. 

Finally, the cavity freely evolves from $t_{K}^{m}$ to the end time
$t_{N}$ (where $t_{N}$ is the same for all $M$ measurement processes),
and the final state reads
\begin{equation}
\vert\psi^{m}(t_{N})\rangle=U(t_{N},t_{K}^{m})\vert\psi^{m}(t_{K}^{m})\rangle=e^{i\Theta_{K+1}^{m}}\vert\Lambda_{\text{f}}^{m}\rangle.\label{eq:psi_T}
\end{equation}
Here, we denote $t_{K+1}^{m}\equiv t_{N}$ and
\begin{equation}
\Lambda_{\text{f}}^{m}=\sum_{k=1}^{K+1}(-1)^{K-k+1}\alpha(t_{k}^{m},t_{k-1}^{m}).
\end{equation}

It is seen from Eq.~(\ref{eq:Lambda_j}) that the CS signal $\Lambda_{\text{f}}^{m}$
is a linear combination of the $K+1$ randomly split segments of the
original signal $\alpha(t,t_{0})$. To express $\Lambda_{\text{f}}^{m}$
with the components of the discretized signal $\vec{\alpha}=(\alpha_{1},\dots,\alpha_{N})^{\text{T}}$,
we define the increase in $\alpha(t,t_{0})$ during a step of evolution
$\tau_{\text{B}}$ by $\beta_{n}\equiv\alpha(n\tau_{\text{B}},(n-1)\tau_{\text{B}})$, and thus $\alpha_{n}=\sum_{l=1}^{n}\beta_{l}$. In this manner, Eq.~(\ref{eq:Lambda_j}) can be rewritten in the form
\begin{equation}
\Lambda_{\text{f}}^{m}=\vec{\mathcal{A}}^{m}\cdot\vec{\beta},\label{eq:Lambda_i2}
\end{equation}
Here, $\vec{\beta}=(\beta_{1},\beta_{2},\dots,\beta_{N})^{\text{T}}$
is an $N$-dimensional column vector and $\vec{\mathcal{A}}^{m}=(\mathcal{A}_{1}^{m},\mathcal{A}_{2}^{m},\dots,\mathcal{A}_{N}^{m})$
is a row vector whose elements are
\begin{equation}
\mathcal{A}_{n}^{m}=\sum_{k=1}^{K+1}(-1)^{K-k+1}H_{k}^{m}(n\tau_{\text{B}}),\label{eq:A_j_i}
\end{equation}
where $n=1,\dots,N,m=1,\dots,M$ and
\begin{equation}
H_{k}^{m}(t)=\begin{cases}
1, & t_{k-1}^{m}<t\leq t_{k}^{m},\\
0, & \text{otherwise}.
\end{cases}\label{eq:H_k_i}
\end{equation}

When all $M$ measurements are complete, the information that we obtained
about $\vec{\alpha}$ is compressed into the vector $\vec{\Lambda}_{\text{f}}=(\Lambda_{\text{f}}^{1},\Lambda_{\text{f}}^{2},\dots,\Lambda_{\text{f}}^{M})^{\text{T}}$.
By arranging the $M$ row vectors $\vec{\mathcal{A}}^{m}$ into an
$M\times N$ dimensional matrix $\mathcal{A}=[\vec{\mathcal{A}}^{1},\vec{\mathcal{A}}^{2},\dots,\vec{\mathcal{A}}^{M}]$, we can express the CS measurement of the cavity field in the form
\begin{equation}
\vec{\Lambda}_{\text{f}}=\mathcal{A}\cdot\vec{\beta}.\label{eq:Lambda}
\end{equation}
A comparison with Eq.~(\ref{eq:CS}) reveals that $\vec{\Lambda}_{\text{f}}$ plays the role of the measurement results $\vec{y}$, $\mathcal{A}$ the sensing matrix, and $\vec{\beta}$ the signal $\vec{x}$. Once we successfully obtain the recovered result $\vec{\beta}^{\prime}$, which should coincide with $\vec{\beta}$ with high accuracy, the coherent field amplitude can be straightforwardly reconstructed using $\alpha_{n}=\sum_{l=1}^{n}\beta_{l}^{\prime}$.

So far, we have systematically introduced the CS approach to measure the
time-dependent cavity field, especially describing how to construct the sensing matrix with coherent operations. It is noted that the methods
used in this cavity QED setup can be generalized to other systems of circuit QED, optomechanics, and nitrogen-vacancy centers for monitoring specific dynamic variables.

\begin{figure*}
\includegraphics[width=8cm]{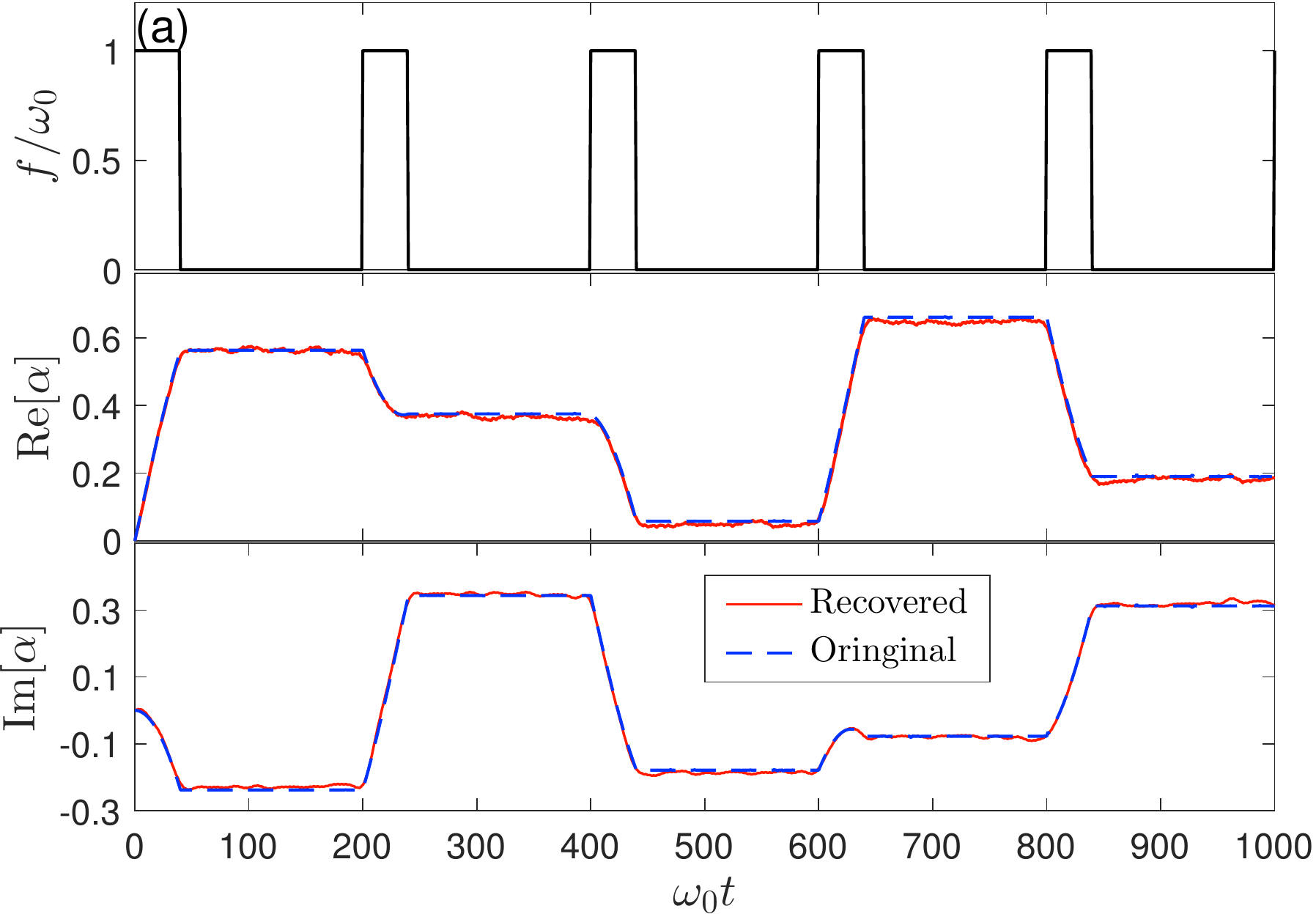}\includegraphics[width=8cm]{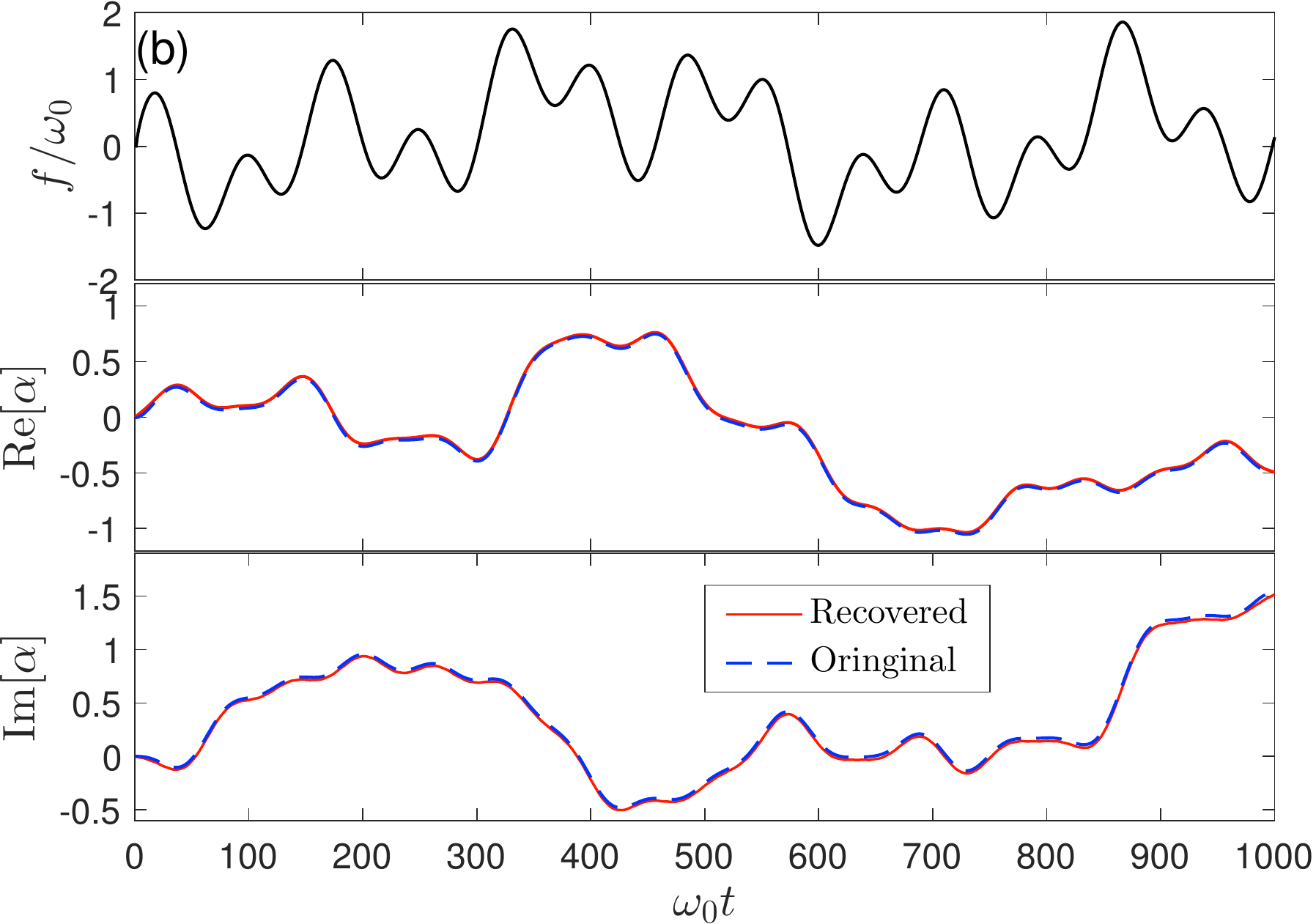}\caption{\label{fig:2}(color online) Recovered amplitudes of the coherent
cavity field driven by (a) periodical squared and (b) randomly generated
protocols. The time-dependent driving fields $f(t)$ are shown in
the upper row. The middle and bottom rows are the real and imaginary
parts of the coherent amplitudes; the red solid lines correspond to
the signals $\vec{\alpha}^{\prime}$ recovered taking the CS approach
in the presence of noise, and the blue dashed lines correspond to
the original signals calculated using Eq.~(\ref{eq:alpha}). The recovered
signals coincide with the original signals with mean-squared error (MSE) less than $5\times10^{-4}$.
We choose the Nyquist--Shannon sampling interval to be $\omega_{0}^{-1}$
and set the total sampling times as $N=1000$. The numbers of CS measurements are $M=220$ and $M=200$ for driving protocols (a) and (b), respectively. The detuning between the driving field and cavity mode frequencies is set as $\Delta=0.02\omega_{0}$.}
\end{figure*}

\section{Recovery results }

In this section, we take two representative driving protocols as examples
to demonstrate the numerical simulation of the signal recovery
via CS techniques in cavity QED setups. In the first protocol,
we design the driving field $f(t)$ as periodical square pulses with
period $\tau=(t_{N}-t_{0})/5$ and a 20\% duty cycle [upper row in
Fig.~\ref{fig:2}(a)]. In the second protocol, the driving field
$f(t)$ is a randomly generated smooth function of time [upper row
in Fig.~\ref{fig:2}(b)]. The original values of the real and imaginary
parts of $\alpha(t,t_{0})$ are calculated using Eq.~(\ref{eq:alpha})
with respect to the two cases and shown by blue dashed lines in the middle and bottom rows of Fig.~\ref{fig:2}, respectively. 

As mentioned in Sec.~\ref{sec:II}, besides dramatically reducing the
sampling number, another advantage of CS is its robustness
against noise. In consideration of practical cases, we add
white noise to the magnitude of the driving strength as $f_{\xi}(t)=f(t)+0.05\omega_{0}\xi$,
where $\xi$ is a stochastic variable uniformly distributed in $[-1,1]$.
In the simulation, we randomly generate an $M\times N$ sensing matrix
$\mathcal{A}$ according to Eq.~(\ref{eq:A_j_i}) and then apply $\mathcal{A}$ to $\vec{\beta}_{\xi}$ in obtaining the sample vector $\vec{\Lambda}_{\text{f}}$. Here, $\vec{\beta}_{\xi}$ is the signal mixed with noise.

To solve the optimization problems defined in Eq.~(\ref{eq:CSrec}), $\vec{\beta}_{\xi}$ should be transformed
into a sparse representation in advance. We assume the signals $\vec{\beta}_{\xi}$
are sparse in the discrete cosine bases, and the corresponding transformation $\Phi$ is expressed by
\[
\Phi_{ij}=\sqrt{\frac{2-\delta_{1,i}}{N}}\cos\left[\frac{\pi}{2N}(i-1)(2j-1)\right],i,j=1,2,\dots,N.
\]
Equation~(\ref{eq:Lambda}) can then be rewritten as $\vec{\Lambda}_{\text{f}}=(\mathcal{A}\Phi^{-1})\cdot(\Phi\vec{\beta}_{\xi})$, where $\Phi\vec{\beta}_{\xi}$ is a sparse vector and $\mathcal{A}\Phi^{-1}$ remains a random matrix. The choice of sparse representation is usually not unique, and the use of different sparse representations (e.g., discrete Fourier bases and Walsh bases) will result in different sparsity of the signal
$\Phi\vec{\beta}_{\xi}$. However, finding the representation with
minimum sparsity is not the goal of this work, and the discrete cosine
bases work well for the signals driven by the two different protocols here.

The total evolution time is fixed and discretized into $N=1000$ uniform
intervals, which is the number of measurements required by the Nyquist--Shannon sampling theorem. In the discrete cosine bases, the sparsity of $\vec{\beta}_{\xi}$
is approximately $S=50$ for the periodic square wave driving protocol
and $S=40$ for the randomly driving protocol. Therefore, the numbers
of measurements needed for the CS approach are estimated as $\sim S\log_{2}(N/S)$ as $M=220$ ($M=200$) in the first (second) protocol. Each row of
the sensing matrix $\mathcal{A}$ comprises $N$ elements of
$\pm1$ with a total of $K$ randomly flips of $1\leftrightarrow-1$. In
the simulation, we set $K=30$ ($K=20$) for the first (second) protocol. 

The real and imaginary parts of $\vec{\beta}_{\xi}$ are separately
recovered from the real and imaginary parts of $\vec{\Lambda}_{\text{f}}$, respectively, via the OMP algorithm. We denote the recovered signal as $\vec{\beta}^{\prime}$, and the coherent field amplitudes $\vec{\alpha}$ are then obtained straightforwardly according to $\alpha_{n}=\sum_{l=1}^{n}\beta_{l}^{\prime}$
and are shown by red solid lines in the middle and bottom rows of
Fig.~\ref{fig:2}, corresponding to the two driving protocols. We see that the recovered field amplitudes coincide well with the theoretically
calculations even in the presence of noise. The MSE of both the real and imaginary parts of $\vec{\alpha}$ are
less than $5\times10^{-4}$.

\begin{figure}
\includegraphics[width=8cm]{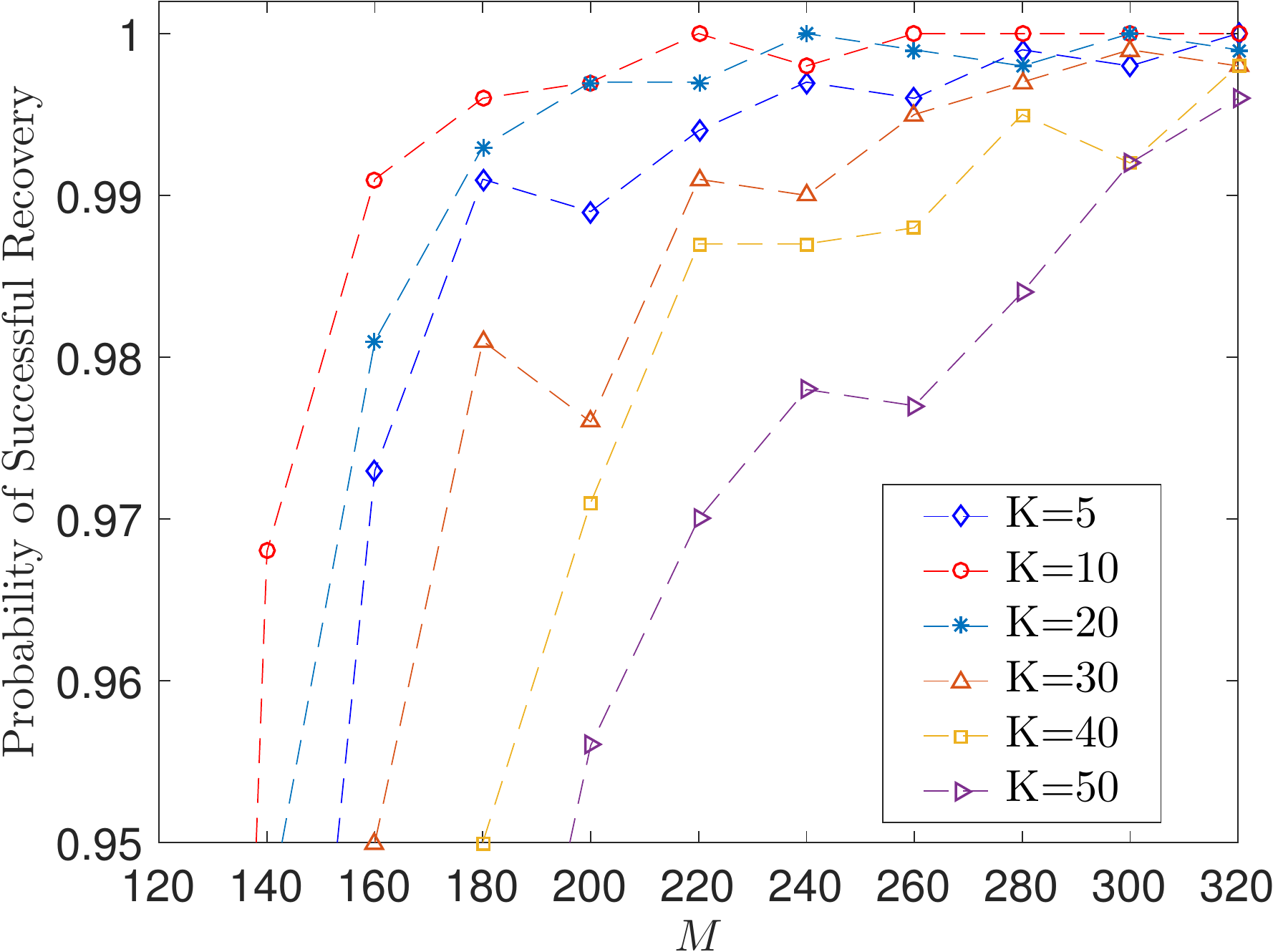}
\caption{\label{fig:3}(color online) Probability of successful recovery
for the randomly driving protocol as a function of $M$. Different
numbers of flips $K$ during a single measurement are presented
with different colored markers.}
\end{figure}

In Fig.~\ref{fig:2}, the number of measurements $M$ is set around
the theoretical lower bound, which is large enough to guarantee the
effectiveness of CS. Nonetheless, as the sensing matrix $\mathcal{A}$
is randomly generated, we cannot guarantee the recovery is successful
every time, even though the failure case is rare. We want $M$ to be
as small as possible meanwhile maintaining good recovery quality. To this
end, we need to check the probability of successful recovery with
respect to different values of $M$ and $K$. For every parameter combination of $\{M,K\}$, we carry out $1000$ simulations and sum the MSEs of the real and imaginary parts of each $\vec{\beta}^{\prime}$. We judge the recovery to be successful if the MSE is less than $2\times10^{-3}$; otherwise, the recovery has failed. The probabilities of successful recovery for the randomly driving protocol are shown in Fig.~\ref{fig:3}. When $M\geq220$, the success probability quickly converges to above 99.5\%, and if we lower the success criterion of the MSE to $5\times10^{-3}$ the success probability stabilizes at 1 for $M\geq180$. 

Another important result obtained from Fig.~\ref{fig:3} is that
$K$ has an approximate optimal value for the success probability
at a fixed $M$. This result can be roughly explained in that each column of $\Phi^{-1}$ is an orthogonal vector of the discrete cosine bases,
and then if the row vectors of $\mathcal{A}$ are almost homogeneously
filled with $1$ or $-1$ (small $K$ case), or almost alternately
filled with $1$ and $-1$ (large $K$ case), then the magnitudes of many
elements of $\mathcal{A}\Phi^{-1}$ will be greatly canceled, which
yields a poor quality sensing matrix $\mathcal{A}\Phi^{-1}$. In the
randomly driving case, the success probability roughly increases
with $K$ when $K$ is less than 10 and then decreases rapidly when $K$
is larger than 20. This favorable result implies that we only need
a small number of atoms to modulate the cavity field in each measurement,
and the error introduced by the imperfect control of the atom--photon
interaction thus does not have an appreciable effect. 

\section{CONCLUSIONS AND OUTLOOK}

We proposed a measurement approach based on CS techniques to monitor
the dynamic evolution of a driven cavity QED system. Taking this approach, the time-varying coherent state of the cavity field can be probed and reconstructed with high accuracy but with an exponentially reduced number of measurements compared with the number required by the Nyquist--Shannon sampling theorem. The key issue is the realization of the random sensing matrix via experimental realizable quantum operations. We suggest that sending a sequence of largely detuned two-level atoms through the cavity at time points indicated by the sensing matrix can modulate the cavity field in the designed CS method. 

The simulation results show that the reconstructed amplitudes of the
coherent cavity field coincide well with the original amplitudes. Even
in the presence of weak white noise, the error can still be suppressed
with MSE no more than $5\times10^{-4}$. Additionally, the dependence of the probability of successful recovery on the numbers of measurements and the number of atoms in a single measurement was investigated, further strengthening the fact that CS techniques can greatly reduce the volume of resources consumed in experiments without affecting the recovery accuracy.

In summary, the quantum measurement method developed in this work reveals the great potential of CS techniques in quantum sensing and metrology, and more efforts should made to incorporate the ingenious concept of CS with the precise quantum control and measurement techniques.
\begin{acknowledgments}
This study is supported by the National Natural Science Foundation
of China (Grant No. 12075025).
\end{acknowledgments}

\bibliographystyle{apsrev4-1}
\bibliography{CS_time_evolution}

\begin{thebibliography}{35}%
\makeatletter
\providecommand \@ifxundefined [1]{%
 \@ifx{#1\undefined}
}%
\providecommand \@ifnum [1]{%
 \ifnum #1\expandafter \@firstoftwo
 \else \expandafter \@secondoftwo
 \fi
}%
\providecommand \@ifx [1]{%
 \ifx #1\expandafter \@firstoftwo
 \else \expandafter \@secondoftwo
 \fi
}%
\providecommand \natexlab [1]{#1}%
\providecommand \enquote  [1]{``#1''}%
\providecommand \bibnamefont  [1]{#1}%
\providecommand \bibfnamefont [1]{#1}%
\providecommand \citenamefont [1]{#1}%
\providecommand \href@noop [0]{\@secondoftwo}%
\providecommand \href [0]{\begingroup \@sanitize@url \@href}%
\providecommand \@href[1]{\@@startlink{#1}\@@href}%
\providecommand \@@href[1]{\endgroup#1\@@endlink}%
\providecommand \@sanitize@url [0]{\catcode `\\12\catcode `\$12\catcode
  `\&12\catcode `\#12\catcode `\^12\catcode `\_12\catcode `\%12\relax}%
\providecommand \@@startlink[1]{}%
\providecommand \@@endlink[0]{}%
\providecommand \url  [0]{\begingroup\@sanitize@url \@url }%
\providecommand \@url [1]{\endgroup\@href {#1}{\urlprefix }}%
\providecommand \urlprefix  [0]{URL }%
\providecommand \Eprint [0]{\href }%
\providecommand \doibase [0]{http://dx.doi.org/}%
\providecommand \selectlanguage [0]{\@gobble}%
\providecommand \bibinfo  [0]{\@secondoftwo}%
\providecommand \bibfield  [0]{\@secondoftwo}%
\providecommand \translation [1]{[#1]}%
\providecommand \BibitemOpen [0]{}%
\providecommand \bibitemStop [0]{}%
\providecommand \bibitemNoStop [0]{.\EOS\space}%
\providecommand \EOS [0]{\spacefactor3000\relax}%
\providecommand \BibitemShut  [1]{\csname bibitem#1\endcsname}%
\let\auto@bib@innerbib\@empty
\bibitem [{\citenamefont {Braginsky}\ \emph {et~al.}(1980)\citenamefont
  {Braginsky}, \citenamefont {Vorontsov},\ and\ \citenamefont
  {Thorne}}]{1980.Thorne}%
  \BibitemOpen
  \bibfield  {author} {\bibinfo {author} {\bibfnamefont {V.~B.}\ \bibnamefont
  {Braginsky}}, \bibinfo {author} {\bibfnamefont {Y.~I.}\ \bibnamefont
  {Vorontsov}}, \ and\ \bibinfo {author} {\bibfnamefont {K.~S.}\ \bibnamefont
  {Thorne}},\ }\href {\doibase 10.1126/science.209.4456.547} {\bibfield
  {journal} {\bibinfo  {journal} {Science}\ }\textbf {\bibinfo {volume}
  {209}},\ \bibinfo {pages} {547} (\bibinfo {year} {1980})}\BibitemShut
  {NoStop}%
\bibitem [{\citenamefont {Misra}\ and\ \citenamefont
  {Sudarshan}(1977)}]{1977.Sudarshan}%
  \BibitemOpen
  \bibfield  {author} {\bibinfo {author} {\bibfnamefont {B.}~\bibnamefont
  {Misra}}\ and\ \bibinfo {author} {\bibfnamefont {E.~C.~G.}\ \bibnamefont
  {Sudarshan}},\ }\href {\doibase 10.1063/1.523304} {\bibfield  {journal}
  {\bibinfo  {journal} {J. Math. Phys.}\ }\textbf {\bibinfo {volume} {18}},\
  \bibinfo {pages} {756} (\bibinfo {year} {1977})}\BibitemShut {NoStop}%
\bibitem [{\citenamefont {Itano}\ \emph {et~al.}(1990)\citenamefont {Itano},
  \citenamefont {Heinzen}, \citenamefont {Bollinger},\ and\ \citenamefont
  {Wineland}}]{1990.Wineland}%
  \BibitemOpen
  \bibfield  {author} {\bibinfo {author} {\bibfnamefont {W.~M.}\ \bibnamefont
  {Itano}}, \bibinfo {author} {\bibfnamefont {D.~J.}\ \bibnamefont {Heinzen}},
  \bibinfo {author} {\bibfnamefont {J.~J.}\ \bibnamefont {Bollinger}}, \ and\
  \bibinfo {author} {\bibfnamefont {D.~J.}\ \bibnamefont {Wineland}},\ }\href
  {\doibase 10.1103/physreva.41.2295} {\bibfield  {journal} {\bibinfo
  {journal} {Phys. Rev. A}\ }\textbf {\bibinfo {volume} {41}},\ \bibinfo
  {pages} {2295} (\bibinfo {year} {1990})}\BibitemShut {NoStop}%
\bibitem [{\citenamefont {Nyquist}(1928)}]{1928.Nyquist}%
  \BibitemOpen
  \bibfield  {author} {\bibinfo {author} {\bibfnamefont {H.}~\bibnamefont
  {Nyquist}},\ }\href {\doibase 10.1109/t-aiee.1928.5055024} {\bibfield
  {journal} {\bibinfo  {journal} {Transactions of the American Institute of
  Electrical Engineers}\ }\textbf {\bibinfo {volume} {47}},\ \bibinfo {pages}
  {617} (\bibinfo {year} {1928})}\BibitemShut {NoStop}%
\bibitem [{\citenamefont {Shannon}(1949)}]{1949.Shannon}%
  \BibitemOpen
  \bibfield  {author} {\bibinfo {author} {\bibfnamefont {C.}~\bibnamefont
  {Shannon}},\ }\href {\doibase 10.1109/jrproc.1949.232969} {\bibfield
  {journal} {\bibinfo  {journal} {Proceedings of the IRE}\ }\textbf {\bibinfo
  {volume} {37}},\ \bibinfo {pages} {10} (\bibinfo {year} {1949})}\BibitemShut
  {NoStop}%
\bibitem [{\citenamefont {Cand\`{e}s}\ \emph {et~al.}(2006)\citenamefont
  {Cand\`{e}s}, \citenamefont {Romberg},\ and\ \citenamefont
  {Tao}}]{2006.Taoitr}%
  \BibitemOpen
  \bibfield  {author} {\bibinfo {author} {\bibfnamefont {E.~J.}\ \bibnamefont
  {Cand\`{e}s}}, \bibinfo {author} {\bibfnamefont {J.}~\bibnamefont {Romberg}},
  \ and\ \bibinfo {author} {\bibfnamefont {T.}~\bibnamefont {Tao}},\ }\href
  {\doibase 10.1109/tit.2005.862083} {\bibfield  {journal} {\bibinfo  {journal}
  {IEEE Trans. Inf. Theory}\ }\textbf {\bibinfo {volume} {52}},\ \bibinfo
  {pages} {489} (\bibinfo {year} {2006})}\BibitemShut {NoStop}%
\bibitem [{\citenamefont {Cand\`{e}s}\ and\ \citenamefont
  {Tao}(2006)}]{2006.Tao}%
  \BibitemOpen
  \bibfield  {author} {\bibinfo {author} {\bibfnamefont {E.~J.}\ \bibnamefont
  {Cand\`{e}s}}\ and\ \bibinfo {author} {\bibfnamefont {T.}~\bibnamefont
  {Tao}},\ }\href {\doibase 10.1109/tit.2006.885507} {\bibfield  {journal}
  {\bibinfo  {journal} {IEEE Trans. Inf. Theory}\ }\textbf {\bibinfo {volume}
  {52}},\ \bibinfo {pages} {5406} (\bibinfo {year} {2006})}\BibitemShut
  {NoStop}%
\bibitem [{\citenamefont {Donoho}(2006)}]{2006.Donoho}%
  \BibitemOpen
  \bibfield  {author} {\bibinfo {author} {\bibfnamefont {D.~L.}\ \bibnamefont
  {Donoho}},\ }\href {\doibase 10.1109/tit.2006.871582} {\bibfield  {journal}
  {\bibinfo  {journal} {IEEE Trans. Inf. Theory}\ }\textbf {\bibinfo {volume}
  {52}},\ \bibinfo {pages} {1289} (\bibinfo {year} {2006})}\BibitemShut
  {NoStop}%
\bibitem [{\citenamefont {Lustig}\ \emph {et~al.}(2007)\citenamefont {Lustig},
  \citenamefont {Donoho},\ and\ \citenamefont {Pauly}}]{2007.Pauly}%
  \BibitemOpen
  \bibfield  {author} {\bibinfo {author} {\bibfnamefont {M.}~\bibnamefont
  {Lustig}}, \bibinfo {author} {\bibfnamefont {D.}~\bibnamefont {Donoho}}, \
  and\ \bibinfo {author} {\bibfnamefont {J.~M.}\ \bibnamefont {Pauly}},\ }\href
  {\doibase 10.1002/mrm.21391} {\bibfield  {journal} {\bibinfo  {journal}
  {Magn. Reson. Med.}\ }\textbf {\bibinfo {volume} {58}},\ \bibinfo {pages}
  {1182} (\bibinfo {year} {2007})}\BibitemShut {NoStop}%
\bibitem [{\citenamefont {Potter}\ \emph {et~al.}(2010)\citenamefont {Potter},
  \citenamefont {Ertin}, \citenamefont {Parker},\ and\ \citenamefont
  {Cetin}}]{2010.Cetin}%
  \BibitemOpen
  \bibfield  {author} {\bibinfo {author} {\bibfnamefont {L.~C.}\ \bibnamefont
  {Potter}}, \bibinfo {author} {\bibfnamefont {E.}~\bibnamefont {Ertin}},
  \bibinfo {author} {\bibfnamefont {J.~T.}\ \bibnamefont {Parker}}, \ and\
  \bibinfo {author} {\bibfnamefont {M.}~\bibnamefont {Cetin}},\ }\href
  {\doibase 10.1109/jproc.2009.2037526} {\bibfield  {journal} {\bibinfo
  {journal} {Proc. IEEE}\ }\textbf {\bibinfo {volume} {98}},\ \bibinfo {pages}
  {1006} (\bibinfo {year} {2010})}\BibitemShut {NoStop}%
\bibitem [{\citenamefont {Liu}\ \emph {et~al.}(2019)\citenamefont {Liu},
  \citenamefont {Yao}, \citenamefont {Liu}, \citenamefont {Xu}, \citenamefont
  {Wang}, \citenamefont {Liu}, \citenamefont {Wang}, \citenamefont {Zhai},\
  and\ \citenamefont {Zhao}}]{2019.Zhao}%
  \BibitemOpen
  \bibfield  {author} {\bibinfo {author} {\bibfnamefont {S.}~\bibnamefont
  {Liu}}, \bibinfo {author} {\bibfnamefont {X.-R.}\ \bibnamefont {Yao}},
  \bibinfo {author} {\bibfnamefont {X.-F.}\ \bibnamefont {Liu}}, \bibinfo
  {author} {\bibfnamefont {D.-Z.}\ \bibnamefont {Xu}}, \bibinfo {author}
  {\bibfnamefont {X.-D.}\ \bibnamefont {Wang}}, \bibinfo {author}
  {\bibfnamefont {B.}~\bibnamefont {Liu}}, \bibinfo {author} {\bibfnamefont
  {C.}~\bibnamefont {Wang}}, \bibinfo {author} {\bibfnamefont {G.-J.}\
  \bibnamefont {Zhai}}, \ and\ \bibinfo {author} {\bibfnamefont
  {Q.}~\bibnamefont {Zhao}},\ }\href {\doibase 10.1364/oe.27.022138} {\bibfield
   {journal} {\bibinfo  {journal} {Opt. Express}\ }\textbf {\bibinfo {volume}
  {27}},\ \bibinfo {pages} {22138} (\bibinfo {year} {2019})}\BibitemShut
  {NoStop}%
\bibitem [{\citenamefont {Lin}\ and\ \citenamefont
  {Herrmann}(2007)}]{2007.Herrmann}%
  \BibitemOpen
  \bibfield  {author} {\bibinfo {author} {\bibfnamefont {T.~T.~Y.}\
  \bibnamefont {Lin}}\ and\ \bibinfo {author} {\bibfnamefont {F.~J.}\
  \bibnamefont {Herrmann}},\ }\href {\doibase 10.1190/1.2750716} {\bibfield
  {journal} {\bibinfo  {journal} {Geophysics}\ }\textbf {\bibinfo {volume}
  {72}},\ \bibinfo {pages} {SM77} (\bibinfo {year} {2007})}\BibitemShut
  {NoStop}%
\bibitem [{\citenamefont {Katz}\ \emph {et~al.}(2009)\citenamefont {Katz},
  \citenamefont {Bromberg},\ and\ \citenamefont
  {Silberberg}}]{2009.Silberberg}%
  \BibitemOpen
  \bibfield  {author} {\bibinfo {author} {\bibfnamefont {O.}~\bibnamefont
  {Katz}}, \bibinfo {author} {\bibfnamefont {Y.}~\bibnamefont {Bromberg}}, \
  and\ \bibinfo {author} {\bibfnamefont {Y.}~\bibnamefont {Silberberg}},\
  }\href {\doibase 10.1063/1.3238296} {\bibfield  {journal} {\bibinfo
  {journal} {Appl. Phys. Lett.}\ }\textbf {\bibinfo {volume} {95}},\ \bibinfo
  {pages} {131110} (\bibinfo {year} {2009})}\BibitemShut {NoStop}%
\bibitem [{\citenamefont {Bromberg}\ \emph {et~al.}(2009)\citenamefont
  {Bromberg}, \citenamefont {Katz},\ and\ \citenamefont
  {Silberberg}}]{2009.Silberberg4a6}%
  \BibitemOpen
  \bibfield  {author} {\bibinfo {author} {\bibfnamefont {Y.}~\bibnamefont
  {Bromberg}}, \bibinfo {author} {\bibfnamefont {O.}~\bibnamefont {Katz}}, \
  and\ \bibinfo {author} {\bibfnamefont {Y.}~\bibnamefont {Silberberg}},\
  }\href {\doibase 10.1103/physreva.79.053840} {\bibfield  {journal} {\bibinfo
  {journal} {Phys. Rev. A}\ }\textbf {\bibinfo {volume} {79}},\ \bibinfo
  {pages} {053840} (\bibinfo {year} {2009})}\BibitemShut {NoStop}%
\bibitem [{\citenamefont {Zerom}\ \emph {et~al.}(2011)\citenamefont {Zerom},
  \citenamefont {Chan}, \citenamefont {Howell},\ and\ \citenamefont
  {Boyd}}]{2011.Boydefx}%
  \BibitemOpen
  \bibfield  {author} {\bibinfo {author} {\bibfnamefont {P.}~\bibnamefont
  {Zerom}}, \bibinfo {author} {\bibfnamefont {K.~W.~C.}\ \bibnamefont {Chan}},
  \bibinfo {author} {\bibfnamefont {J.~C.}\ \bibnamefont {Howell}}, \ and\
  \bibinfo {author} {\bibfnamefont {R.~W.}\ \bibnamefont {Boyd}},\ }\href
  {\doibase 10.1103/physreva.84.061804} {\bibfield  {journal} {\bibinfo
  {journal} {Phys. Rev. A}\ }\textbf {\bibinfo {volume} {84}},\ \bibinfo
  {pages} {061804} (\bibinfo {year} {2011})}\BibitemShut {NoStop}%
\bibitem [{\citenamefont {Gross}\ \emph {et~al.}(2010)\citenamefont {Gross},
  \citenamefont {Liu}, \citenamefont {Flammia}, \citenamefont {Becker},\ and\
  \citenamefont {Eisert}}]{2010.Eisert4qn}%
  \BibitemOpen
  \bibfield  {author} {\bibinfo {author} {\bibfnamefont {D.}~\bibnamefont
  {Gross}}, \bibinfo {author} {\bibfnamefont {Y.-K.}\ \bibnamefont {Liu}},
  \bibinfo {author} {\bibfnamefont {S.~T.}\ \bibnamefont {Flammia}}, \bibinfo
  {author} {\bibfnamefont {S.}~\bibnamefont {Becker}}, \ and\ \bibinfo {author}
  {\bibfnamefont {J.}~\bibnamefont {Eisert}},\ }\href {\doibase
  10.1103/physrevlett.105.150401} {\bibfield  {journal} {\bibinfo  {journal}
  {Phys. Rev. Lett.}\ }\textbf {\bibinfo {volume} {105}},\ \bibinfo {pages}
  {150401} (\bibinfo {year} {2010})}\BibitemShut {NoStop}%
\bibitem [{\citenamefont {Cramer}\ \emph {et~al.}(2010)\citenamefont {Cramer},
  \citenamefont {Plenio}, \citenamefont {Flammia}, \citenamefont {Somma},
  \citenamefont {Gross}, \citenamefont {Bartlett}, \citenamefont
  {Landon-Cardinal}, \citenamefont {Poulin},\ and\ \citenamefont
  {Liu}}]{2010.Liu}%
  \BibitemOpen
  \bibfield  {author} {\bibinfo {author} {\bibfnamefont {M.}~\bibnamefont
  {Cramer}}, \bibinfo {author} {\bibfnamefont {M.~B.}\ \bibnamefont {Plenio}},
  \bibinfo {author} {\bibfnamefont {S.~T.}\ \bibnamefont {Flammia}}, \bibinfo
  {author} {\bibfnamefont {R.}~\bibnamefont {Somma}}, \bibinfo {author}
  {\bibfnamefont {D.}~\bibnamefont {Gross}}, \bibinfo {author} {\bibfnamefont
  {S.~D.}\ \bibnamefont {Bartlett}}, \bibinfo {author} {\bibfnamefont
  {O.}~\bibnamefont {Landon-Cardinal}}, \bibinfo {author} {\bibfnamefont
  {D.}~\bibnamefont {Poulin}}, \ and\ \bibinfo {author} {\bibfnamefont {Y.-K.}\
  \bibnamefont {Liu}},\ }\href {\doibase 10.1038/ncomms1147} {\bibfield
  {journal} {\bibinfo  {journal} {Nat. Commun.}\ }\textbf {\bibinfo {volume}
  {1}},\ \bibinfo {pages} {149} (\bibinfo {year} {2010})}\BibitemShut {NoStop}%
\bibitem [{\citenamefont {Shabani}\ \emph {et~al.}(2011)\citenamefont
  {Shabani}, \citenamefont {Kosut}, \citenamefont {Mohseni}, \citenamefont
  {Rabitz}, \citenamefont {Broome}, \citenamefont {Almeida}, \citenamefont
  {Fedrizzi},\ and\ \citenamefont {White}}]{2011.White}%
  \BibitemOpen
  \bibfield  {author} {\bibinfo {author} {\bibfnamefont {A.}~\bibnamefont
  {Shabani}}, \bibinfo {author} {\bibfnamefont {R.~L.}\ \bibnamefont {Kosut}},
  \bibinfo {author} {\bibfnamefont {M.}~\bibnamefont {Mohseni}}, \bibinfo
  {author} {\bibfnamefont {H.}~\bibnamefont {Rabitz}}, \bibinfo {author}
  {\bibfnamefont {M.~A.}\ \bibnamefont {Broome}}, \bibinfo {author}
  {\bibfnamefont {M.~P.}\ \bibnamefont {Almeida}}, \bibinfo {author}
  {\bibfnamefont {A.}~\bibnamefont {Fedrizzi}}, \ and\ \bibinfo {author}
  {\bibfnamefont {A.~G.}\ \bibnamefont {White}},\ }\href {\doibase
  10.1103/physrevlett.106.100401} {\bibfield  {journal} {\bibinfo  {journal}
  {Phys. Rev. Lett.}\ }\textbf {\bibinfo {volume} {106}},\ \bibinfo {pages}
  {100401} (\bibinfo {year} {2011})}\BibitemShut {NoStop}%
\bibitem [{\citenamefont {Rudinger}\ and\ \citenamefont
  {Joynt}(2015)}]{2015.Joynt}%
  \BibitemOpen
  \bibfield  {author} {\bibinfo {author} {\bibfnamefont {K.}~\bibnamefont
  {Rudinger}}\ and\ \bibinfo {author} {\bibfnamefont {R.}~\bibnamefont
  {Joynt}},\ }\href {\doibase 10.1103/physreva.92.052322} {\bibfield  {journal}
  {\bibinfo  {journal} {Phys. Rev. A}\ }\textbf {\bibinfo {volume} {92}},\
  \bibinfo {pages} {052322} (\bibinfo {year} {2015})}\BibitemShut {NoStop}%
\bibitem [{\citenamefont {Riofr\'{i}o}\ \emph {et~al.}(2017)\citenamefont
  {Riofr\'{i}o}, \citenamefont {Gross}, \citenamefont {Flammia}, \citenamefont
  {Monz}, \citenamefont {Nigg}, \citenamefont {Blatt},\ and\ \citenamefont
  {Eisert}}]{2017.Eiserth38}%
  \BibitemOpen
  \bibfield  {author} {\bibinfo {author} {\bibfnamefont {C.~A.}\ \bibnamefont
  {Riofr\'{i}o}}, \bibinfo {author} {\bibfnamefont {D.}~\bibnamefont {Gross}},
  \bibinfo {author} {\bibfnamefont {S.~T.}\ \bibnamefont {Flammia}}, \bibinfo
  {author} {\bibfnamefont {T.}~\bibnamefont {Monz}}, \bibinfo {author}
  {\bibfnamefont {D.}~\bibnamefont {Nigg}}, \bibinfo {author} {\bibfnamefont
  {R.}~\bibnamefont {Blatt}}, \ and\ \bibinfo {author} {\bibfnamefont
  {J.}~\bibnamefont {Eisert}},\ }\href {\doibase 10.1038/ncomms15305}
  {\bibfield  {journal} {\bibinfo  {journal} {Nat. Commun.}\ }\textbf {\bibinfo
  {volume} {8}},\ \bibinfo {pages} {15305} (\bibinfo {year}
  {2017})}\BibitemShut {NoStop}%
\bibitem [{\citenamefont {Dunbar}\ \emph {et~al.}(2013)\citenamefont {Dunbar},
  \citenamefont {Osborne}, \citenamefont {Anna},\ and\ \citenamefont
  {Kubarych}}]{2013.Kubarych}%
  \BibitemOpen
  \bibfield  {author} {\bibinfo {author} {\bibfnamefont {J.~A.}\ \bibnamefont
  {Dunbar}}, \bibinfo {author} {\bibfnamefont {D.~G.}\ \bibnamefont {Osborne}},
  \bibinfo {author} {\bibfnamefont {J.~M.}\ \bibnamefont {Anna}}, \ and\
  \bibinfo {author} {\bibfnamefont {K.~J.}\ \bibnamefont {Kubarych}},\ }\href
  {\doibase 10.1021/jz401281r} {\bibfield  {journal} {\bibinfo  {journal} {J.
  Phys. Chem. Lett.}\ }\textbf {\bibinfo {volume} {4}},\ \bibinfo {pages}
  {2489} (\bibinfo {year} {2013})}\BibitemShut {NoStop}%
\bibitem [{\citenamefont {Roeding}\ \emph {et~al.}(2017)\citenamefont
  {Roeding}, \citenamefont {Klimovich},\ and\ \citenamefont
  {Brixner}}]{2017.Brixner}%
  \BibitemOpen
  \bibfield  {author} {\bibinfo {author} {\bibfnamefont {S.}~\bibnamefont
  {Roeding}}, \bibinfo {author} {\bibfnamefont {N.}~\bibnamefont {Klimovich}},
  \ and\ \bibinfo {author} {\bibfnamefont {T.}~\bibnamefont {Brixner}},\ }\href
  {\doibase 10.1063/1.4976309} {\bibfield  {journal} {\bibinfo  {journal} {J.
  Chem. Phys.}\ }\textbf {\bibinfo {volume} {146}},\ \bibinfo {pages} {084201}
  (\bibinfo {year} {2017})}\BibitemShut {NoStop}%
\bibitem [{\citenamefont {Magesan}\ \emph {et~al.}(2013)\citenamefont
  {Magesan}, \citenamefont {Cooper},\ and\ \citenamefont
  {Cappellaro}}]{2013.Cappellaro}%
  \BibitemOpen
  \bibfield  {author} {\bibinfo {author} {\bibfnamefont {E.}~\bibnamefont
  {Magesan}}, \bibinfo {author} {\bibfnamefont {A.}~\bibnamefont {Cooper}}, \
  and\ \bibinfo {author} {\bibfnamefont {P.}~\bibnamefont {Cappellaro}},\
  }\href {\doibase 10.1103/physreva.88.062109} {\bibfield  {journal} {\bibinfo
  {journal} {Phys. Rev. A}\ }\textbf {\bibinfo {volume} {88}},\ \bibinfo
  {pages} {062109} (\bibinfo {year} {2013})}\BibitemShut {NoStop}%
\bibitem [{\citenamefont {Ladd}\ \emph {et~al.}(2010)\citenamefont {Ladd},
  \citenamefont {Jelezko}, \citenamefont {Laflamme}, \citenamefont {Nakamura},
  \citenamefont {Monroe},\ and\ \citenamefont {O'Brien}}]{2010.O'Brien}%
  \BibitemOpen
  \bibfield  {author} {\bibinfo {author} {\bibfnamefont {T.~D.}\ \bibnamefont
  {Ladd}}, \bibinfo {author} {\bibfnamefont {F.}~\bibnamefont {Jelezko}},
  \bibinfo {author} {\bibfnamefont {R.}~\bibnamefont {Laflamme}}, \bibinfo
  {author} {\bibfnamefont {Y.}~\bibnamefont {Nakamura}}, \bibinfo {author}
  {\bibfnamefont {C.}~\bibnamefont {Monroe}}, \ and\ \bibinfo {author}
  {\bibfnamefont {J.~L.}\ \bibnamefont {O'Brien}},\ }\href {\doibase
  10.1038/nature08812} {\bibfield  {journal} {\bibinfo  {journal} {Nature}\
  }\textbf {\bibinfo {volume} {464}},\ \bibinfo {pages} {45} (\bibinfo {year}
  {2010})}\BibitemShut {NoStop}%
\bibitem [{\citenamefont {Malnou}\ \emph {et~al.}(2019)\citenamefont {Malnou},
  \citenamefont {Palken}, \citenamefont {Brubaker}, \citenamefont {Vale},
  \citenamefont {Hilton},\ and\ \citenamefont {Lehnert}}]{2019.Lehnert}%
  \BibitemOpen
  \bibfield  {author} {\bibinfo {author} {\bibfnamefont {M.}~\bibnamefont
  {Malnou}}, \bibinfo {author} {\bibfnamefont {D.~A.}\ \bibnamefont {Palken}},
  \bibinfo {author} {\bibfnamefont {B.~M.}\ \bibnamefont {Brubaker}}, \bibinfo
  {author} {\bibfnamefont {L.~R.}\ \bibnamefont {Vale}}, \bibinfo {author}
  {\bibfnamefont {G.~C.}\ \bibnamefont {Hilton}}, \ and\ \bibinfo {author}
  {\bibfnamefont {K.~W.}\ \bibnamefont {Lehnert}},\ }\href {\doibase
  10.1103/physrevx.9.021023} {\bibfield  {journal} {\bibinfo  {journal} {Phys.
  Rev. X}\ }\textbf {\bibinfo {volume} {9}},\ \bibinfo {pages} {021023}
  (\bibinfo {year} {2019})}\BibitemShut {NoStop}%
\bibitem [{\citenamefont {Lewis-Swan}\ \emph {et~al.}(2020)\citenamefont
  {Lewis-Swan}, \citenamefont {Barberena}, \citenamefont {Muniz}, \citenamefont
  {Cline}, \citenamefont {Young}, \citenamefont {Thompson},\ and\ \citenamefont
  {Rey}}]{2020.Rey}%
  \BibitemOpen
  \bibfield  {author} {\bibinfo {author} {\bibfnamefont {R.~J.}\ \bibnamefont
  {Lewis-Swan}}, \bibinfo {author} {\bibfnamefont {D.}~\bibnamefont
  {Barberena}}, \bibinfo {author} {\bibfnamefont {J.~A.}\ \bibnamefont
  {Muniz}}, \bibinfo {author} {\bibfnamefont {J.~R.~K.}\ \bibnamefont {Cline}},
  \bibinfo {author} {\bibfnamefont {D.}~\bibnamefont {Young}}, \bibinfo
  {author} {\bibfnamefont {J.~K.}\ \bibnamefont {Thompson}}, \ and\ \bibinfo
  {author} {\bibfnamefont {A.~M.}\ \bibnamefont {Rey}},\ }\href {\doibase
  10.1103/physrevlett.124.193602} {\bibfield  {journal} {\bibinfo  {journal}
  {Phys. Rev. Lett.}\ }\textbf {\bibinfo {volume} {124}},\ \bibinfo {pages}
  {193602} (\bibinfo {year} {2020})}\BibitemShut {NoStop}%
\bibitem [{\citenamefont {Baraniuk}\ \emph {et~al.}(2008)\citenamefont
  {Baraniuk}, \citenamefont {Davenport}, \citenamefont {DeVore},\ and\
  \citenamefont {Wakin}}]{2008.Wakinfgl}%
  \BibitemOpen
  \bibfield  {author} {\bibinfo {author} {\bibfnamefont {R.}~\bibnamefont
  {Baraniuk}}, \bibinfo {author} {\bibfnamefont {M.}~\bibnamefont {Davenport}},
  \bibinfo {author} {\bibfnamefont {R.}~\bibnamefont {DeVore}}, \ and\ \bibinfo
  {author} {\bibfnamefont {M.}~\bibnamefont {Wakin}},\ }\href {\doibase
  10.1007/s00365-007-9003-x} {\bibfield  {journal} {\bibinfo  {journal}
  {Constr. Approx.}\ }\textbf {\bibinfo {volume} {28}},\ \bibinfo {pages} {253}
  (\bibinfo {year} {2008})}\BibitemShut {NoStop}%
\bibitem [{\citenamefont {Pati}\ \emph {et~al.}(1993)\citenamefont {Pati},
  \citenamefont {Rezaiifar},\ and\ \citenamefont
  {Krishnaprasad}}]{1993.Krishnaprasad}%
  \BibitemOpen
  \bibfield  {author} {\bibinfo {author} {\bibfnamefont {Y.~C.}\ \bibnamefont
  {Pati}}, \bibinfo {author} {\bibfnamefont {R.}~\bibnamefont {Rezaiifar}}, \
  and\ \bibinfo {author} {\bibfnamefont {P.~S.}\ \bibnamefont
  {Krishnaprasad}},\ }\href@noop {} {\bibfield  {journal} {\bibinfo  {journal}
  {In Proc. of the 27th Asilomar Conference on Signals, Systems and Computers}\
  }\textbf {\bibinfo {volume} {1}},\ \bibinfo {pages} {40} (\bibinfo {year}
  {1993})}\BibitemShut {NoStop}%
\bibitem [{\citenamefont {Lvovsky}\ and\ \citenamefont
  {Raymer}(2009)}]{2009.Raymer}%
  \BibitemOpen
  \bibfield  {author} {\bibinfo {author} {\bibfnamefont {A.~I.}\ \bibnamefont
  {Lvovsky}}\ and\ \bibinfo {author} {\bibfnamefont {M.~G.}\ \bibnamefont
  {Raymer}},\ }\href {\doibase 10.1103/revmodphys.81.299} {\bibfield  {journal}
  {\bibinfo  {journal} {Rev. Mod. Phys.}\ }\textbf {\bibinfo {volume} {81}},\
  \bibinfo {pages} {299} (\bibinfo {year} {2009})}\BibitemShut {NoStop}%
\bibitem [{\citenamefont {Brune}\ \emph {et~al.}(1992)\citenamefont {Brune},
  \citenamefont {Haroche}, \citenamefont {Raimond}, \citenamefont
  {Davidovich},\ and\ \citenamefont {Zagury}}]{1992.Zagury}%
  \BibitemOpen
  \bibfield  {author} {\bibinfo {author} {\bibfnamefont {M.}~\bibnamefont
  {Brune}}, \bibinfo {author} {\bibfnamefont {S.}~\bibnamefont {Haroche}},
  \bibinfo {author} {\bibfnamefont {J.~M.}\ \bibnamefont {Raimond}}, \bibinfo
  {author} {\bibfnamefont {L.}~\bibnamefont {Davidovich}}, \ and\ \bibinfo
  {author} {\bibfnamefont {N.}~\bibnamefont {Zagury}},\ }\href {\doibase
  10.1103/physreva.45.5193} {\bibfield  {journal} {\bibinfo  {journal} {Phys.
  Rev. Lett.}\ }\textbf {\bibinfo {volume} {45}},\ \bibinfo {pages} {5193}
  (\bibinfo {year} {1992})}\BibitemShut {NoStop}%
\bibitem [{\citenamefont {Nogues}\ \emph {et~al.}(1999)\citenamefont {Nogues},
  \citenamefont {Rauschenbeutel}, \citenamefont {Osnaghi}, \citenamefont
  {Brune}, \citenamefont {Raimond},\ and\ \citenamefont
  {Haroche}}]{1999.Haroche}%
  \BibitemOpen
  \bibfield  {author} {\bibinfo {author} {\bibfnamefont {G.}~\bibnamefont
  {Nogues}}, \bibinfo {author} {\bibfnamefont {A.}~\bibnamefont
  {Rauschenbeutel}}, \bibinfo {author} {\bibfnamefont {S.}~\bibnamefont
  {Osnaghi}}, \bibinfo {author} {\bibfnamefont {M.}~\bibnamefont {Brune}},
  \bibinfo {author} {\bibfnamefont {J.~M.}\ \bibnamefont {Raimond}}, \ and\
  \bibinfo {author} {\bibfnamefont {S.}~\bibnamefont {Haroche}},\ }\href
  {\doibase 10.1038/22275} {\bibfield  {journal} {\bibinfo  {journal} {Nature}\
  }\textbf {\bibinfo {volume} {400}},\ \bibinfo {pages} {239} (\bibinfo {year}
  {1999})}\BibitemShut {NoStop}%
\bibitem [{\citenamefont {Hansen}\ \emph {et~al.}(2001)\citenamefont {Hansen},
  \citenamefont {Aichele}, \citenamefont {Hettich}, \citenamefont {Lodahl},
  \citenamefont {Lvovsky}, \citenamefont {Mlynek},\ and\ \citenamefont
  {Schiller}}]{2001.Schiller}%
  \BibitemOpen
  \bibfield  {author} {\bibinfo {author} {\bibfnamefont {H.}~\bibnamefont
  {Hansen}}, \bibinfo {author} {\bibfnamefont {T.}~\bibnamefont {Aichele}},
  \bibinfo {author} {\bibfnamefont {C.}~\bibnamefont {Hettich}}, \bibinfo
  {author} {\bibfnamefont {P.}~\bibnamefont {Lodahl}}, \bibinfo {author}
  {\bibfnamefont {A.~I.}\ \bibnamefont {Lvovsky}}, \bibinfo {author}
  {\bibfnamefont {J.}~\bibnamefont {Mlynek}}, \ and\ \bibinfo {author}
  {\bibfnamefont {S.}~\bibnamefont {Schiller}},\ }\href {\doibase
  10.1364/ol.26.001714} {\bibfield  {journal} {\bibinfo  {journal} {Opt.
  Lett.}\ }\textbf {\bibinfo {volume} {26}},\ \bibinfo {pages} {1714} (\bibinfo
  {year} {2001})}\BibitemShut {NoStop}%
\bibitem [{\citenamefont {Ueda}\ \emph {et~al.}(1992)\citenamefont {Ueda},
  \citenamefont {Imoto}, \citenamefont {Nagaoka},\ and\ \citenamefont
  {Ogawa}}]{1992.Ogawa}%
  \BibitemOpen
  \bibfield  {author} {\bibinfo {author} {\bibfnamefont {M.}~\bibnamefont
  {Ueda}}, \bibinfo {author} {\bibfnamefont {N.}~\bibnamefont {Imoto}},
  \bibinfo {author} {\bibfnamefont {H.}~\bibnamefont {Nagaoka}}, \ and\
  \bibinfo {author} {\bibfnamefont {T.}~\bibnamefont {Ogawa}},\ }\href
  {\doibase 10.1103/physreva.46.2859} {\bibfield  {journal} {\bibinfo
  {journal} {Phys. Rev. A}\ }\textbf {\bibinfo {volume} {46}},\ \bibinfo
  {pages} {2859} (\bibinfo {year} {1992})}\BibitemShut {NoStop}%
\bibitem [{\citenamefont {Bernu}\ \emph {et~al.}(2008)\citenamefont {Bernu},
  \citenamefont {Del\'{e}glise}, \citenamefont {Sayrin}, \citenamefont {Kuhr},
  \citenamefont {Dotsenko}, \citenamefont {Brune}, \citenamefont {Raimond},\
  and\ \citenamefont {Haroche}}]{2008.Haroche}%
  \BibitemOpen
  \bibfield  {author} {\bibinfo {author} {\bibfnamefont {J.}~\bibnamefont
  {Bernu}}, \bibinfo {author} {\bibfnamefont {S.}~\bibnamefont
  {Del\'{e}glise}}, \bibinfo {author} {\bibfnamefont {C.}~\bibnamefont
  {Sayrin}}, \bibinfo {author} {\bibfnamefont {S.}~\bibnamefont {Kuhr}},
  \bibinfo {author} {\bibfnamefont {I.}~\bibnamefont {Dotsenko}}, \bibinfo
  {author} {\bibfnamefont {M.}~\bibnamefont {Brune}}, \bibinfo {author}
  {\bibfnamefont {J.~M.}\ \bibnamefont {Raimond}}, \ and\ \bibinfo {author}
  {\bibfnamefont {S.}~\bibnamefont {Haroche}},\ }\href {\doibase
  10.1103/physrevlett.101.180402} {\bibfield  {journal} {\bibinfo  {journal}
  {Phys. Rev. Lett.}\ }\textbf {\bibinfo {volume} {101}},\ \bibinfo {pages}
  {180402} (\bibinfo {year} {2008})}\BibitemShut {NoStop}%
\bibitem [{\citenamefont {Xu}\ \emph {et~al.}(2011)\citenamefont {Xu},
  \citenamefont {Ai},\ and\ \citenamefont {Sun}}]{2011.Sun}%
  \BibitemOpen
  \bibfield  {author} {\bibinfo {author} {\bibfnamefont {D.~Z.}\ \bibnamefont
  {Xu}}, \bibinfo {author} {\bibfnamefont {Q.}~\bibnamefont {Ai}}, \ and\
  \bibinfo {author} {\bibfnamefont {C.~P.}\ \bibnamefont {Sun}},\ }\href
  {\doibase 10.1103/physreva.83.022107} {\bibfield  {journal} {\bibinfo
  {journal} {Phys. Rev. A}\ }\textbf {\bibinfo {volume} {83}},\ \bibinfo
  {pages} {022107} (\bibinfo {year} {2011})}\BibitemShut {NoStop}%
\end{thebibliography}%

\end{document}